\documentclass[conference,letterpaper]{IEEEtran}

\ifCLASSINFOpdf
\else
\fi
\usepackage{float}
\usepackage{graphicx}
\usepackage{subfigure}
\usepackage{multicol}
\usepackage{algorithm}
\usepackage{algpseudocode}
\usepackage{amsmath}
\usepackage{amssymb}
\usepackage{mathrsfs}
\usepackage{amsthm}
\usepackage{bm}
\usepackage{cite}
\usepackage{enumerate}
\usepackage{diagbox}
\usepackage{color}

\newcommand{\cM}{\mathcal{M}}
\newcommand{\cK}{\mathcal{K}}
\newcommand{\cP}{\mathcal{P}}
\newcommand{\cN}{\mathcal{N}}

\newcommand{\cG}{\mathcal{G}}
\newcommand{\cH}{\mathcal{H}}

\newcommand{\cU}{\mathcal{U}}
\newcommand{\cD}{\mathcal{D}}
\newcommand{\cI}{\mathcal{I}}

\newcommand{\bw}{\boldsymbol w}
\newcommand{\bp}{\boldsymbol p}

\newcommand{\bR}{\boldsymbol R}

\newtheorem{myth}{Theorem}
\newtheorem{mydef}[myth]{Definition} 

\newtheorem{myle}[myth]{Lemma}

\IEEEoverridecommandlockouts

\begin{document}
\title{Optimal Joint Subcarrier and Power Allocation in NOMA is Strongly NP-Hard}

\author{
\thanks{A part of the work was carried out at LINCS (www.lincs.fr).}%
\IEEEauthorblockN{Lou Sala\"un\IEEEauthorrefmark{1}\IEEEauthorrefmark{2}, Chung~Shue~Chen\IEEEauthorrefmark{1} and Marceau~Coupechoux\IEEEauthorrefmark{2}}
\IEEEauthorblockA{\IEEEauthorrefmark{1}Bell Labs, Nokia Paris-Saclay, 91620 Nozay, France}
\IEEEauthorblockA{\IEEEauthorrefmark{2}LTCI, Telecom ParisTech, University Paris-Saclay, France}
\IEEEauthorblockA{Email: lou.salaun@nokia-bell-labs.com,  chung\_shue.chen@nokia-bell-labs.com, marceau.coupechoux@telecom-paristech.fr}
}

\maketitle

\begin{abstract}
Non-orthogonal multiple access (NOMA) is a promising radio access technology for 5G. It allows several users to transmit on the same frequency and time resource by performing power-domain multiplexing. At the receiver side, successive interference cancellation (SIC) is applied to mitigate interference among the multiplexed signals. In this way, NOMA can outperform orthogonal multiple access schemes used in conventional cellular networks in terms of spectral efficiency and allows more simultaneous users. This paper investigates the computational complexity of joint subcarrier and power allocation problems in multi-carrier NOMA systems. We prove that these problems are strongly NP-hard for a large class of objective functions, namely the weighted generalized means of the individual data rates. This class covers the popular weighted sum-rate, proportional fairness, harmonic mean and max-min fairness utilities. Our results show that the optimal power and subcarrier allocation cannot be computed in polynomial time in the general case, unless P = NP. Nevertheless, we present some tractable special cases and we show that they can be solved efficiently.
\end{abstract}


\section{Introduction}
Long Term Evolution 4G standards have adopted orthogonal multiple access (OMA) schemes for downlink (OFDMA)~\cite{dahlman20134g} as well as for uplink (SC-FDMA)~\cite{myung2006single}. OMA schemes aim to avoid or alleviate mutual interference among the users by dividing the radio resource into interference-free blocks. While this strategy allows low-complexity signal decoding at the receiver side, its spectral efficiency is suboptimal~\cite{cover2012elements} due to the orthogonal channel access requirement. 

The fifth generation (5G) mobile networks is facing new challenges. Some key requirements are high data rates, improved spectral efficiency and massive device connectivity. Non-orthogonal multiple access (NOMA) is a promising technology to meet these requirements, and has recently received significant attention~\cite{dai2015non}. In contrast to OMA, NOMA allows to multiplex several users on the same radio resource block, therefore achieving higher system spectral efficiency~\cite{saito2013non}. Realistic system-level simulations in~\cite{benjebbour2013system} demonstrate that NOMA achieves higher throughput than OFDMA in downlink. Besides,~\cite{ding2014performance} shows through analytical results that NOMA can achieve superior ergodic sum-rate performance.

In multi-carrier systems, the total bandwidth is divided into subcarriers. The basic principle of multi-carrier NOMA (MC-NOMA) is to superpose several users' signals on the same subcarrier and to perform successive interference cancellation (SIC) at the receiver side to mitigate the co-channel interference. Power allocation among multiplexed users of the same subcarrier should be optimized to achieve desirable data rate performance. Since the number of superposed signals per subcarrier should be limited due to error propagation and decoding complexity concerns in practice~\cite{tse2005fundamentals}, it is also important to optimize the subcarrier allocation for the users.

Several papers in the literature have developed algorithms for joint subcarrier and power allocation with the aim of maximizing some system utility functions such as the sum of data rates, proportional fairness and max-min fairness. Fractional transmit power control (FTPC) is commonly used for sum-rate maximization~\cite{saito2013non,ding2016impact}, it allocates fraction of the total power budget to each user based on their respective channel condition. In~\cite{chen2015suboptimal} and~\cite{al2014uplink}, heuristic user pairing strategies and iterative resource allocation algorithms were studied for uplink transmissions. Reference~\cite{parida2014power} developed a greedy user selection and sub-optimal power allocation scheme based on difference-of-convex programming to maximize the weighted sum-rate. User selection algorithms for sum-rate and proportional fairness utilities were studied in~\cite{datta2016optimal}. The authors of~\cite{lei2016power} solved the downlink sum-rate maximization problem by a Lagrangian duality and dynamic programming algorithm, and derived an upper bound for the achievable sum-rate. This algorithm is mostly used as a benchmark due to its high computational complexity. A more efficient heuristic based on iterative waterfilling method is introduced in~\cite{fu2017double}.

The aforementioned papers have proposed heuristic schemes to solve various difficult joint subcarrier and power allocation problems in NOMA. In order to understand how well can these problems be solved in practical systems with limited computational resources, it is important to study their computational complexity. Moreover, knowing the complexity due to different objective functions and system constraints would complete our understanding on how to design NOMA algorithms. However, to the best of our knowledge, only few papers have studied these problems from a computational complexity point of view:~\cite{lei2016power} proved that sum-rate maximization in downlink MC-NOMA is strongly NP-hard, whereas~\cite{liu2014complexity2,liu2014complexity1,hayashi2009spectrum} proved that several problems involving various utilities and constraints are strongly NP-hard in OFDMA systems. The latter results can be seen as special cases of MC-NOMA with only one user transmitting on each subcarrier. 

Motivated by the above observation, we propose to study the computational complexity of a large class of joint subcarrier and power allocation problems in MC-NOMA systems. We aim at developing a framework to cover most problems introduced in the literature, while also taking into account practical constraints. The complexity analysis is provided to have a more complete understanding of NOMA optimization problems and to facilitate resource allocation algorithm design. Besides, several results and techniques used in the derivation could be interesting or reused for similar problems. More precisely, the contributions of this paper are: 
\begin{enumerate}
\item We prove that utility maximization problems in MC-NOMA are strongly NP-hard for a large class of objective functions, namely the weighted generalized means of order $i\leq 1$ of the individual data rates. This class covers the popular weighted sum-rate, proportional fairness, harmonic mean and max-min fairness utilities. 
\item We develop a unique framework to study these problems based on polynomial reductions from the 3-dimensional matching~\cite{karp1972reducibility}. Our result includes both downlink and uplink cases. It also takes into account individual power budget constraints as well as superposed coding and SIC practical limitations. 
\item This class of problems is a general extension of the existing joint subcarrier and power allocation problems studied in the literature. Indeed, previous papers~\cite{liu2014complexity2,liu2014complexity1,hayashi2009spectrum} focused on OFDMA, while~\cite{lei2016power} considered only sum-rate maximization in downlink MC-NOMA, which are both special cases of our framework. 
\item Finally, we show some interesting special cases of our problem which are solvable in polynomial time. In addition, we provide a short discussion on possible algorithms to solve them. 
\end{enumerate}

The rest of the paper is organized as follows. In Section~\ref{sec:model}, we present the system model and notations. In Section~\ref{sec:prob}, we formulate the class of utility maximization problems to be studied. By computational complexity analysis, we prove in Section~\ref{sec:cplx} that these problems are strongly NP-hard. Some tractable special cases are also presented. Finally, we conclude in Section~\ref{sec:ccl}.

\section{System Model}\label{sec:model}
We consider a multi-carrier NOMA system with one base station (BS) serving $K$ users on $N$ subcarriers. Let ${\cK \triangleq \{1,\ldots,K\}}$ denotes the index set of all users, and ${\cN \triangleq \{1,\ldots,N\}}$ the index set of all subcarriers. For ${n\in\cN}$, let $W_n$ be the bandwidth of subcarrier $n$ and we use ${W = \sum_{n\in\cN}{W_n}}$ to denote the total bandwidth. We consider there is no interference between adjacent subcarriers due to orthogonal frequency division. Moreover, we assume that each subcarrier $n\in\cN$ experiences frequency-flat block fading on its bandwidth $W_n$.

For $k\in\cK$ and $n\in\cN$, let $g_k^n$ and $\eta_k^n$ be the link gain and the received noise power of user $k$ on subcarrier $n$. 
We assume that the channel can be accessed through either downlink or uplink transmissions. For downlink scenarios, the BS transmits a signal to each user $k$ on subcarrier $n$ with power $p_k^n$. For uplink scenarios, each user $k$ transmits a signal to the BS on subcarrier $n$ with power $p_k^n$. In both cases, we refer to $p_k^n$ as ``the allocated transmit power of user $k$'' on subcarrier $n$.
User $k$ is said to be active on subcarrier $n$ if $p_k^n>0$. 

We consider a power domain NOMA system in which up to $M$ users can be multiplexed on the same subcarrier using superposition coding. Variable $M$ is a parameter of the system, and its value depends on practical limitations of SIC due to decoding complexity and error propagation~\cite{tse2005fundamentals}.
Let ${\cU_n \triangleq \{k\in\cK \colon p_k^n >0\}}$ represents the set of users multiplexed on subcarrier $n$. Each subcarrier is modeled as a multi-user Gaussian broadcast channel~\cite{cover2012elements,tse2005fundamentals} and SIC is applied at the receiver side to mitigate intra-band interference.

In order to model SIC, we need to consider the decoding order of every user $k\in\cU_n$ multiplexed on the same subcarrier $n\in\cN$. This decoding order is represented by a permutation function $\pi_n \colon \{1,\ldots,|\cU_n|\} \to \cU_n$, where $|\mathord{\cdot}|$ denotes the cardinality of a finite set. For $i\in\{1,\ldots,|\cU_n|\}$, $\pi_n(i)$ returns the $i$-th decoded user's index. Conversely, user $k$'s decoding order is given by $\pi_n^{-1}(k)$. Hence, the signals of users ${\pi_n(1),\ldots,\pi_n(i-1)}$ are first decoded and subtracted from the superposed signal before decoding $\pi_n(i)$'s signal. Furthermore, user $\pi_n(i)$ is subject to interference from users $\pi_n(j)$, for $j > i$. In particular, $\pi_n(|\cU_n|)$ is decoded last and is not subject to any intra-band interference if the previous $|\cU_n|-1$ users have been successfully decoded. 

Note that the above discussion can be applied to both uplink and downlink cases. However, the decoding order should be chosen differently depending on which case we are addressing. For downlink scenarios, the optimal decoding order obeys the following sorting~\cite[Section 6.2]{tse2005fundamentals}:
\begin{equation}\label{SIC_DL}
\frac{\eta_{\pi_n(1)}^n}{g_{\pi_n(1)}^n} \geq \frac{\eta_{\pi_n(2)}^n}{g_{\pi_n(2)}^n} \geq \cdots \geq \frac{\eta_{\pi_n(|\cU_n|)}^n}{g_{\pi_n(|\cU_n|)}^n}.
\end{equation}
For uplink scenarios, the decoding starts from the strongest user first and move towards the weakest user~\cite[Section 6.1]{tse2005fundamentals}:
\begin{equation}\label{SIC_UL}
g_{\pi_n(1)}^n p_{\pi_n(1)}^n  \geq g_{\pi_n(2)}^n p_{\pi_n(2)}^n \geq \cdots \geq g_{\pi_n(|\cU_n|)}^n p_{\pi_n(|\cU_n|)}^n.
\end{equation}
It is worth mentioning that the complexity results in the paper can be easily adapted to any chosen decoding order. Therefore, without loss of generality, we will only consider the above~(\ref{SIC_DL}) and~(\ref{SIC_UL}) decoding orders for downlink and uplink scenarios, respectively. 

Shannon capacity formula is applied to model the capacity of a communication link, i.e., the maximum achievable data rate. Regarding the downlink, the achievable data rate of user $k\in\cK$ on subcarrier $n\in\cN$ is given by:
\begin{equation*} \label{R_DL}
R_k^n \triangleq W_n\log_2\left(1+\frac{g_k^n p_k^n}{\sum_{j=\pi^{-1}_n(k)+1}^{|\cU_n|}{g_k^n p_{\pi_n(j)}^n}+\eta^n_k}\right).
\end{equation*}
In the uplink, since the only receiver is the BS, all users transmitting on subcarrier $n$ are subject to the same noise $\eta^n = \eta_k^n$, $k\in\cK$. The data rate is then expressible as:
\begin{equation*} \label{R_UL}
R_k^n \triangleq W_n\log_2\left(1+\frac{g_k^n p_k^n}{\sum_{j=\pi^{-1}_n(k)+1}^{|\cU_n|}{g_{\pi_n(j)}^n p_{\pi_n(j)}^n}+\eta^n}\right).
\end{equation*}
In accordance with the SIC decoding order, user $k$ is only subject to interference from users $\pi_n(j)$, ${j>\pi^{-1}_n(k)}$. 

For ease of reading, let us define the following notations: $R_k \triangleq \sum_{n\in\cN}{R_k^n}$ represents user $k$'s individual data rate, while $R^n \triangleq \sum_{k\in\cK}{R_k^n}$ corresponds to the sum of data rates achieved on subcarrier $n$. We denote by ${\bp \triangleq\left(p_k^n\right)_{k\in\cK,n\in\cN}}$ and ${\bR \triangleq \left(R_k\right)_{k\in\cK}}$ the power allocation vector and individual data rates vector, respectively. Data rates are function of the power allocation, nevertheless we use the notations $R_k^n$, $R_k$, $R^n$ and $\bR$ instead of $R_k^n(\bp)$, $R_k(\bp)$, $R^n(\bp)$, $\bR(\bp)$, for simplicity.

\section{Problem Formulation}\label{sec:prob}
\begin{mydef}[Weighted generalized mean~\cite{hardy1952inequalities}]
$ $\newline
Let $\cM_{i,\bw}$ denotes the weighted generalized mean of order $i\in\mathbb{R}\setminus\{0\}$, which is defined with a sequence of positive weights ${\bw = \{w_1,\ldots,w_K\}}$ such that $\sum_{k=1}^K{w_k}=1$. For $K$ positive real numbers $x_1,\ldots,x_K$, we have:
\begin{equation*}\label{eqn:mean1}
\cM_{i,\bw}\left(x_1,\ldots,x_K\right) = \left(\sum_{k=1}^K{w_k x_k^i}\right)^{1/i}.
\end{equation*}
It can also be extended to $i\in\{-\infty,0\}$ by taking the limit, i.e., $\cM_{i,\bw}\left(x_1,\ldots,x_K\right) = \lim_{j \rightarrow i} \cM_{j,\bw}\left(x_1,\ldots,x_K\right)$. 

An important property is the generalized mean inequality, see below:
\begin{equation}\label{proof:NPhard_mean_ineq}
r < q \implies \cM_{r,\bw}\left(x_1,\ldots,x_K\right) \leq \cM_{q,\bw}\left(x_1,\ldots,x_K\right).
\end{equation}
Note that the equality holds if and only if (iff) ${x_1=\cdots=x_K}$.
\end{mydef}

In this work, we will focus on the following class of joint subcarrier and power allocation problems for any ${i\in\left[-\infty,1\right]}$:
\begin{equation}\tag{$\cP_i$}\label{P}
\begin{aligned}
& \underset{\bp}{\text{maximize}}
& & \cM_{i,\bw}\left(\bR\left(\bp\right)\right), \\
& \text{subject to}
& & C1:~\sum_{n\in\cN}p_k^n \leq \bar{P}_k,~k\in\cK, \\
&&& C2:~p_k^n \leq \bar{p}_k^n,~k\in\cK,~n\in\cN, \\
&&& C3:~p_k^n \geq 0,~k\in\cK,~n\in\cN, \\
&&& C4:~|\cU_n| \leq M,~n\in\cN. \\
\end{aligned}
\end{equation}
Note that $C1$ represents each user's total power budget. $C2$ is the power constraint on each subcarrier. $C3$ ensures that the allocated powers remain non-negative. Due to decoding complexity and error propagation in SIC~\cite{tse2005fundamentals}, practical implementation has a maximum number of multiplexed users per subcarrier $M$, which corresponds to constraint $C4$.

The objective of~\ref{P} is to maximize the weighted generalized mean value of every individual data rates $R_k$, $k\in\cK$. This objective function is also known as $\alpha$-fairness~\cite{mo2000fair,altman2008generalized}. We can see that when $\bw =\{1/K,\ldots,1/K\}$, we have the following popular utility functions:
\begin{enumerate}
\item Sum rate utility, namely the arithmetic mean: 
\begin{equation*} 
\cM_{1,\bw} = \frac{1}{K}\sum_{k=1}^K{R_k}
\end{equation*}
\item Proportional fairness utility, namely the geometric mean:
\begin{equation*} 
\cM_{0,\bw} = \left(\prod_{k=1}^K{R_k}\right)^{1/K}
\end{equation*}
\item Harmonic mean utility:
\begin{equation*} 
\cM_{-1,\bw} = K/\left(\sum_{k=1}^K{\left(R_k\right)^{-1}}\right)
\end{equation*}
\item Max-min utility:
\begin{equation*} 
\cM_{-\infty,\bw} = \min_{k\in\cK}\{R_k\}
\end{equation*}
\end{enumerate}
Note that most utility maximization problems in the NOMA literature belong to the above four utilities~\cite{lei2016power,ding2014performance,benjebbour2013system,ding2016impact,chen2015suboptimal,parida2014power,al2014uplink,datta2016optimal,fu2017double,tse2005fundamentals}. However, our study of the general problem~\ref{P} could provide a foundation applicable to a much larger scope of similar subjects.

For the sake of completeness, Definition~\ref{def:NPO} formalizes the idea of NP optimization problems introduced in~\cite{Hochbaum}.
\begin{mydef}[NP optimization problem (NPO)]\label{def:NPO}
$ $\newline
A NPO problem $\cH$ is a 4-tuple $\left(\mathcal{I}_\cH,\mathcal{S}_\cH,f,\text{type}\right)$ such that
\begin{enumerate}
\item $\mathcal{I}_\cH$ is the set of instances. Each instance is recognizable in polynomial time.
\item For any instance $x\in\mathcal{I}_\cH$, $\mathcal{S}_\cH(x)$ is the space of feasible solutions. Every solution $y\in\mathcal{S}_\cH(x)$ has a size bounded by a polynomial in the size of $x$. Moreover, membership in $\mathcal{S}_\cH$ is decidable in polynomial time.
\item $f$ is the objective function, computable in polynomial time.
\item $\text{type}\in\{\text{min},\text{max}\}$ indicates whether $\cH$ is a minimization or maximization problem.
\end{enumerate}
\end{mydef}

Notice that $\cP_i = \left(\mathcal{I}_{\cP_i},\mathcal{S}_{\cP_i},\cM_{i,\bw},\text{max}\right)$ is a NP optimization problem since it fulfills Definition~\ref{def:NPO}. $\mathcal{I}_{\cP_i}$ contains all the system parameters presented in Section~\ref{sec:model}, i.e.,
\begin{multline}\label{param}
\mathcal{I}_{\cP_i} = (\bw,K,N,M,(W_n)_{n\in\cN},(g_k^n)_{n\in\cN,k\in\cK},\\ 
(\eta_k^n)_{n\in\cN,k\in\cK},(\bar{P}_k)_{k\in\cK},(\bar{p}_k^n)_{n\in\cN,k\in\cK}).
\end{multline}
For a given instance $x\in\mathcal{I}_{\cP_i}$, the feasible set $\mathcal{S}_{\cP_i}(x)$ is defined as the set of all power vectors satisfying constraints $C1$ to $C4$. Condition \textit{2)} in Definition~\ref{def:NPO} holds, since these constraints can be verified in polynomial time. Finally, as required by condition \textit{3)}, the objective function is computable in polynomial time.

\section{Computational Complexity}\label{sec:cplx}
\subsection{Definitions and Preliminaries}
Let $opt_{\cP_i}(x)$ be the global optimal of an instance $x\in\mathcal{I}_{\cP_i}$, then the decision version of problem~\ref{P} consists of checking if this value is greater or equal to a given threshold $T$, i.e.,
\begin{equation}\tag{${\cD_i}$}\label{P_D}
opt_{\cP_i}(x) \geq T.
\end{equation}
In Garey and Johnson computational complexity framework~\cite[Chapter 5]{garey2002computers}, a numerical optimization problem \ref{P} is said to be NP-hard if its corresponding decision problem~\ref{P_D} is NP-hard. A discussion about strong NP-hardness and complexity preserving reductions can be found in~\cite{garey1978strong}. We summarize these concepts in Definition~\ref{def:NPhard}. 

\begin{mydef}[Strong NP-hardness]\label{def:NPhard}
$ $\newline
A decision problem $\cH$ is said to be NP-hard if there exists a polynomial-time reduction from a NP-complete problem $\cG$ to $\cH$. In addition, $\cH$ is said to be strongly NP-hard if it is still NP-hard even when all its numerical parameters are bounded by a polynomial in the size of the input. 
\end{mydef}

The 3-Dimensional Matching Problem (3DM) is one of Karp's 21 NP-complete problems~\cite{karp1972reducibility} and is also known to be NP-complete in the strong sense~\cite{garey1975complexity}. 

\begin{mydef}[3-Dimensional Matching Problem]\label{def:3DM}
$ $\newline
The 3-dimensional matching problem (3DM) takes four finite sets as inputs $(X,Y,Z,S)$ such that $|X|=|Y|=|Z|$ and $S \subseteq X \times Y \times Z$. Let $\cI_{3DM}$ denotes the set of all possible inputs. The problem consists of deciding whether there exists a \mbox{3-dimensional matching} $S' \subseteq S$ such that no two distinct triplets $(x_1,y_1,z_1),\, (x_2,y_2,z_2)\in S'$ overlap, i.e.,
\begin{equation*}\label{def:3DM_1}
(x_1,y_1,z_1) \neq (x_2,y_2,z_2) \implies x_1 \neq x_2, y_1\neq y_2, z_1\neq z_2,
\end{equation*}
and all elements are covered by $S'$, i.e.,
\begin{equation}\label{def:3DM_2}
|S'| = |X|.
\end{equation}
\end{mydef}

The subclass of problems~\ref{P} and~\ref{P_D} in which the number of multiplexed users per subcarrier $M$ is fixed is denoted by $\cP_{i|M}$ and $\cD_{i|M}$, respectively. Any instance of the optimization problem can be converted to an instance of the decision problem by appending an additional threshold parameter $T\in\mathbb{R}$,
\begin{equation}\label{D_instances}
\cI_{\cD_{i|M}} = \cI_{\cP_{i|M}} \times \mathbb{R}.
\end{equation}

In this work, we consider pseudo-polynomial reductions $t_M \colon \cI_{3DM} \mapsto \cI_{\cD_{i|M}}$ mapping any instance of 3DM to an instance of ${\cD_{i|M}}$, for $M \geq 1$, $i\in\left[-\infty,1\right]$. Pseudo-polynomial transformations preserve NP-hardness in the strong sense~\cite{garey1978strong} and are defined for all instances $x_{3DM}\in\cI_{3DM}$ as:
\begin{enumerate}[(i)]
\item $x_{3DM}$ has a matching $\iff opt_{\cP_i}(t_M(x_{3DM})) \geq T$,
\item $t_M$ is polynomial time computable in the size of $x_{3DM}$,
\item The largest numerical value of $t_M(x_{3DM})$ is lower and upper bounded by polynomials in the size of $x_{3DM}$.
\end{enumerate}

In the following subsections, we will prove that $\cD_{i|M}$ is strongly NP-hard for any fixed $M\geq 1$ and ${i\in\left[-\infty,1\right]}$ by constructing the aforementioned pseudo-polynomial reduction $t_M$. To this end, we first prove it in Lemma~\ref{le:NPhard1} for the sum-rate objective function $\cM_{1,\bw}$ with no more than $M=1$ user multiplexed per subcarrier. Then, we extend this proof in Lemma~\ref{le:NPhard2} to any $M\geq 1$. In Theorem~\ref{th:NPhard}, we generalize it to any objective functions $\cM_{i,\bw}$, $i \leq 1$. It is interesting to note that OFDMA results~\cite{liu2014complexity2,liu2014complexity1,hayashi2009spectrum} correspond to the special case $M=1$ of Theorem~\ref{th:NPhard}. Finally, we discuss some special cases solvable in polynomial time.
\subsection{Sum-Rate Maximization with $M=1$}\label{subsec:NPhard1}
\begin{myle}\label{le:NPhard1}
For $M = 1$, problem~${\cD_{1|M}}$ with sum-rate objective function $\cM_{1,\bw}$ is strongly NP-hard in both downlink and uplink scenarios.
\end{myle}
\begin{IEEEproof}
The idea of the proof is to construct a reduction $t_1 \colon \cI_{3DM} \mapsto \cI_{\cD_{1|1}}$ mapping any instance of 3DM to an instance of ${\cD_{1|1}}$ in which no more than one user is allocated to each subcarrier. We first detail $t_1$ and show that it satisfies conditions (ii) and (iii). Then, we prove condition (i) for $T=3$. As a result, $t_1$ is a well defined pseudo-polynomial reduction, and it follows from 3DM's strong NP-hardness~\cite{garey1975complexity} that~${\cD_{1|1}}$ is also strongly NP-hard.

Let $x_{3DM}=(X,Y,Z,S)\in\cI_{3DM}$. Without loss of generality, we can assume that $|S| \geq |X|$, otherwise $x_{3DM}$ has trivially no matching according to~(\ref{def:3DM_2}). The corresponding instance $t_1(x_{3DM})\in\cI_{\cD_{1|1}}$ is given by:
\begin{itemize}
\item $K=|S|$ users. There is a bijective mapping between users $k\in\cK$ and triplets $(x_k,y_k,z_k)\in S$.
\item $N = |S|+2|X|$ subcarriers divided into four groups $\cN_X$, $\cN_Y$, $\cN_Z$ and $\cN_R$.
\end{itemize}
The first three groups $\cN_X$, $\cN_Y$, $\cN_Z$ are called \textit{primary} subcarriers and are in bijection with $X$, $Y$ and $Z$ respectively. For notational simplicity, we index them by their corresponding set, e.g., $n_x\in \cN_X$ corresponds to $x\in X$. The same goes for $Y$ and $Z$. This way, we have ${N_X=N_Y=N_Z=|X|}$ subcarriers in each of these primary groups. The set $\cN_R$ is called the \textit{residual} group, it contains $N_R = |S|-|X|$ subcarriers. \\
The link gains of user $k\in\cK$ whose corresponding triplet is $(x_k,y_k,z_k)\in S$ are set as follows:
\begin{equation*}\label{proof:g}
\forall n\in\cN,\; g_k^n = \begin{cases}
             1  & \text{if } n\in\{n_{x_k},n_{y_k},n_{z_k}\},\\
             1  & \text{if } n\in\cN_R, \\
             0  & \text{otherwise}.
       \end{cases}
\end{equation*}
And the noise powers are set as follows:
\begin{equation*}\label{proof:eta}
\forall n\in\cN,\; \eta_k^n = \begin{cases}
             3/7  & \text{if } n\in\cN_R, \\
             1  & \text{otherwise}.
       \end{cases}
\end{equation*}
Noise powers are the same for all users on a given subcarrier, therefore both downlink and uplink scenarios are covered in this proof. We further consider equal weights $\bw =\{1/K,\ldots,1/K\}$ in the objective function, assume that $W_n = 1$ for all ${n\in\cN}$ and $\bar{P}_k = 3$ for all $k\in\cK$. Moreover, its allocated power on each subcarrier $n\in\cN$ is subject to constraint $C2$ such that:
\begin{equation}\label{proof:P_n}
\bar{p}_k^n = \begin{cases}
             1  & \text{if } n\in\cN_X\cup\cN_Y\cup\cN_Z,\\
             3  & \text{if } n\in\cN_R.
       \end{cases}
\end{equation}

For this reduction, we set the decision problem's threshold~(\ref{D_instances}) to be $T=3$. We have characterized above the transformed instance $t_1(x_{3DM})$ and its parameters. The number of parameters is polynomially bounded in the size of $x$: there are $|S|$ users, $|S|+2|X|$ subcarriers, and so on. Thus, by construction our reduction satisfies property (ii). Condition (iii) is also satisfied, since all numerical values are constant, regardless of the size of $x$. It only remains to prove (i) in order to conclude that $t_1$ is indeed a pseudo-polynomial reduction, i.e.,
\begin{equation}\label{proof:goal}
x_{3DM} \text{ has a matching} \iff opt_{\cP_1}(t_1(x_{3DM})) \geq 3.
\end{equation}
No more than one user can be served on each subcarrier according to ${M=1}$ in $C4$. If we suppose that the total system power $\sum_{k\in\cK}{\bar{P}_k} = 3K$ can be distributed among the $N$ subcarriers without constraint $C1$, then the optimal is obtained by the following waterfilling power allocation~\cite{cover2012elements}:
\begin{align}
&\forall n\in\cN_R,\; R^n = \log_2(1+\frac{3}{3/7}) = 3, \label{proof:up1}\\
&\forall n\in\cN_X\cup\cN_Y\cup\cN_Z,\; R^n = \log_2(1+\frac{1}{1}) = 1. \label{proof:up2}
\end{align}
The best solution consists in having the maximum allowable power on every subcarrier while meeting the constraint $C2$. The corresponding user allocation allocates one user on every primary subcarrier with maximum power 1 and one user per residual subcarrier with maximum power 3. According to the problem setting, there is no other optimal power and subcarrier allocation.
There are $3|X|$ primary subcarriers and $|S|-|X|$ residual subcarriers, thus the sum-rate objective is:
\begin{equation}\label{proof:up}
\cM_{1,\bw}\left(\bR\right) = \frac{3|X|\times 1 + (|S|-|X|)\times 3}{K} = 3.
\end{equation}
Since our problem is constrained by $C1$, the optimal cannot be greater than~(\ref{proof:up}), i.e., $opt_{\cP_1}(t_1(x_{3DM})) \leq 3$. It follows that the equivalence~(\ref{proof:goal}) to prove together with the derived upper bound can be rewritten as:
\begin{equation}\label{proof:equality}
x_{3DM} \text{ has a matching} \iff opt_{\cP_1}(t_1(x_{3DM})) = 3.
\end{equation}

Proof of the part $\impliedby$: Let $x_{3DM}=(X, Y, Z, S)$ be an instance of 3DM. Assume that the corresponding instance $t_1(x_{3DM})$ has a power and subcarrier allocation which is optimal and equal to $3$. We have seen that the only possibility to achieve this optimum is to allocate every triplet of subcarriers ${(n_x, n_y, n_z) \in X\times Y\times Z}$ to a user for which channel gain is $1$ with power $1$ and every residual subcarrier to the remaining $|S|-|X|$ users with power $3$. Now, let us define $S'\subset S$ such that $(x,y,z)\in S'$ iff $n_x$, $n_y$ and $n_z$ are allocated to the same user. By construction, $S'$ is a matching for $x_{3DM}$. 

Proof of the part $\implies$: Let $x_{3DM}=(X, Y, Z, S)$ be an instance of 3DM for which there exists a matching $S'$. Consider the following power and subcarrier allocation: for every indexes $(x_k,y_k,z_k) \in S'$ allocate subcarriers $n_{x_k}$, $n_{y_k}$, $n_{z_k}$ to user $k$ with power $1$; allocate the remaining users to the residual subcarriers with power $3$. Then the objective function is exactly $3$, which is also an upper bound. As a consequence, $opt(t_1(x_{3DM}))=3$.  
\end{IEEEproof}

\subsection{Sum-Rate Maximization with $M\geq 1$}\label{subsec:NPhard2}
\begin{myle}\label{le:NPhard2}
For any $M \geq 1$, problem~${\cD_{1|M}}$ with sum-rate objective function $\cM_{1,\bw}$ is strongly NP-hard in both downlink and uplink scenarios.
\end{myle}
\begin{IEEEproof}
The idea of this proof is to extend Lemma~\ref{le:NPhard1}'s reduction to any $M \geq 1$ by adding $N(M-1)$ dummy users. For each subcarrier $n\in\cN$, we create $M-1$ dummy users, denoted by the index set $\cD^n=\{d_1^n,\ldots,d_{M-1}^n\}$. Thus, the set of all users becomes $\cK' = \cK \cup \cD^1 \cup \cdots \cup \cD^N$, where $\cK$ is the users set defined in Lemma~\ref{le:NPhard1}'s proof. All parameters of the transformation $t_M(x_{3DM})\in\cI_{\cD_{1|M}}$ related to user $k\in\cK$ and subcarriers $n\in\cN$ remain as $t_1(x_{3DM})$ in Lemma~\ref{le:NPhard1}'s proof. In addition, we keep equal weights $\bw =\{1/|\cK'|,\ldots,1/|\cK'|\}$. \\
The following construction aims to guarantee that dummy users in $\cD^n$ can only be active on subcarrier $n$. For any ${j\in\{1,\ldots,M-1\}}$, parameters of user ${d_j^n}$ on subcarrier ${n'\in\cN}$ are set as follows:
\begin{equation}\label{proof:dummy_g}
g_{d_j^n}^{n'} = \begin{cases}
             1  & \text{if } n'=n,\\
             0  & \text{otherwise}.
       \end{cases}
\end{equation}
and, 
\begin{equation}
\bar{p}_{d_j^n}^{n'} = \begin{cases}
				\bar{P}_{d_j^n} & \text{if } n'=n,\\
        0  & \text{otherwise}.
				\end{cases}
\end{equation} 
The total power constraint $C1$ is extended as follows:
\begin{equation}\label{proof:dummy_P}
\bar{P}_{d_j^n} = \begin{cases} 
				14 \times 8^{M-j-1} & \text{if } n\in\cN_X\cup\cN_Y\cup\cN_Z,\\
				24 \times 8^{M-j-1} & \text{if } n\in\cN_R.
				\end{cases}
\end{equation} 

Let $\bp^*=\left(p_k^{n*}\right)_{k\in\cK',n\in\cN}$ denotes the optimal power allocation of $t_M(x_{3DM})$. Let $n\in\cN$, since dummy users in $\cD^n$ have greater power budget than any other user in $\cK$ (compare~(\ref{proof:dummy_P}) to~(\ref{proof:P_n})), it is straightforward to see that the optimal is achieved when all $M-1$ dummy users in $\cD^n$ are multiplexed on subcarrier $n$ with the following power allocation:
\begin{equation}\label{proof:dummy_power}
\forall j\in\{1,\ldots,M-1\},\; p_{d_{j}^n}^{n*} = \bar{P}_{d_j^n}.
\end{equation} 
We consider the following decoding order:
\begin{equation}\label{proof:order}
\forall j\in\{1,\ldots,M-1\},\; \pi_n(j) = d_{j}^n.
\end{equation} 
This decoding order satisfies~(\ref{SIC_DL}) and~(\ref{SIC_UL}), therefore both downlink and uplink scenarios are covered in this proof. It is interesting to note that any desired decoding order can be achieved by adjusting the above dummy users' link gains, noise powers and power budgets. 

It remains that subcarrier $n$ can be allocated to an additional non-dummy user $k\in\cK$, while respecting constraints $C4$. In this case, according to~(\ref{SIC_DL}) and~(\ref{SIC_UL}), user $k$ is decoded last, i.e., $\pi_n(M) = k$. Thus, $k$ is not subject to interference from the dummy users on subcarrier $n$. Furthermore, at the optimal, no more than one user in $\cK$ can be multiplexed on each subcarrier $n$. It follows that the optimal subcarrier and power allocation of users $\cK$ in $t_M(x_{3DM})$ is the same as in $t_1(x_{3DM})$ and we have:
\begin{multline}\label{proof:equ_M_1}
opt_{\cP_1}(t_1(x_{3DM})) = 3 \iff\\
opt_{\cP_1}(t_M(x_{3DM})) = \frac{3K+\sum_{n\in\cN}\sum_{j=1}^{M-1}{R_{d_j^n}(\bp^*)}}{|\cK'|}.
\end{multline}
Using~(\ref{proof:dummy_g}-\ref{proof:order}), we can compute the optimal data rate of dummy user ${d_j^n}$, for any ${j\in\{1,\ldots,M-1\}}$, on primary subcarriers $n\in\cN_X\cup\cN_Y\cup\cN_Z$ as
\begin{align}
R_{d_j^n} &= \log_2(1+\frac{\bar{P}_{d_j^n}}{\sum_{j'=j+1}^{M-1}{\bar{P}_{d_j^n}}+2}) \nonumber\\
&= \log_2(1+\frac{14 \times 8^{M-j-1}}{\sum_{j'=j+1}^{M-1}{14 \times 8^{M-j'-1}}+2}) \nonumber\\
&= \log_2(1+\frac{14 \times 8^{M-j-1}}{14(1-8^{M-j-1})/(1-8)+2}) \label{proof:comp_R_primary_1}\\
&= 3 \label{proof:comp_R_primary}
\end{align}
where~(\ref{proof:comp_R_primary_1}) is obtained by calculating the partial sum of the geometric sequence $\sum_{j'=j+1}^{M-1}{14 \times 8^{M-j'-1}}$ with ratio $8$ and $M-j-1$ terms. In the same way, we prove that for all residual subcarriers $n\in\cN_R$, 
\begin{equation}\label{proof:comp_R_residual}
R_{d_j^n} = 3.
\end{equation}
Combining~(\ref{proof:comp_R_primary}) and~(\ref{proof:comp_R_residual}), equivalence~(\ref{proof:equ_M_1}) then becomes 
\begin{align}\label{proof:equ_M_2}
&opt_{\cP_1}(t_1(x_{3DM})) = 3 \nonumber\\
&\iff opt_{\cP_1}(t_M(x_{3DM})) = \frac{3K+3N(M-1)}{|\cK'|} = 3.
\end{align}
Last equality is deduced from ${|\cK'|} = {|\cK \cup \cD^1 \cup \cdots \cup \cD^N|} = {K + N(M-1)}$. Equivalence~(\ref{proof:equality2}) follows from~(\ref{proof:equ_M_2}) and~(\ref{proof:equality}), which implies that $t_M$ is a pseudo-polynomial reduction.
\begin{equation}\label{proof:equality2}
x_{3DM} \text{ has a matching} \iff opt_{\cP_1}(t_M(x_{3DM})) = 3.
\end{equation}
We then conclude from~(\ref{proof:equality2}) and Lemma~\ref{le:NPhard1} that~${\cD_{1|M}}$ is also strongly NP-hard, for any $M\geq 1$.
\end{IEEEproof}

\subsection{Generalized Mean Utility Maximization with $M\geq 1$}\label{subsec:NPhard}
\begin{myth}\label{th:NPhard}
For any ${i\in\left[-\infty,1\right]}$ and $M \geq 1$, problem~${\cD_{i|M}}$ with objective function $\cM_{i,\bw}$ is strongly NP-hard in both downlink and uplink scenarios. In particular, the sum-rate $\cM_{1,\bw}$, proportional fairness $\cM_{0,\bw}$, harmonic mean utility $\cM_{-1,\bw}$ and max-min fairness $\cM_{-\infty,\bw}$ versions of the problem are all strongly NP-hard. 
\end{myth}
\begin{IEEEproof}
Let ${i\in\left[-\infty,1\right)}$, $M \geq 1$ and $x_{3DM}\in\cI_{3DM}$ be an instance of 3DM.
Using Lemma~\ref{le:NPhard2}'s reduction $t_M$, we showed that finding a 3-dimensional matching of $x_{3DM}$ is equivalent to verifying $opt_{\cP_1}(t_M(x_{3DM})) = 3$, i.e.,~(\ref{proof:equality2}). More precisely, when~(\ref{proof:equality2}) is satisfied, all users achieve the same data rate. Indeed, for any user $k\in\cK$, there are three possibilities:
\begin{itemize}
\item $k$ is a dummy user then ${R_k = 3}$ according to~(\ref{proof:comp_R_primary}) and~(\ref{proof:comp_R_residual}), or
\item $k$ is not a dummy user and it is active on a residual subcarrier $n\in\cN_R$ with power $3$ so that ${R_k} = {R_k^n} = {\log_2(1+\frac{3}{3/7})} = 3$, i.e.,~(\ref{proof:up1}), or
\item $k$ is not a dummy user and it is active on three primary subcarriers $n_{x_k}\in\cN_X$, $n_{y_k}\in\cN_Y$ and $n_{z_k}\in\cN_Z$ so that $R_k = R_k^{n_{x_k}}+R_k^{n_{y_k}}+R_k^{n_{z_k}} = 3\log_2(1+\frac{1}{1}) = 3$, i.e.,~(\ref{proof:up2}).
\end{itemize}
It follows that:
\begin{equation}
x_{3DM} \text{ has a matching} \iff \forall k\in\cK, \; R_k(\bp^*) = 3, \label{proof:NPhard_2}
\end{equation}
where $\bp^*$ is the optimal power allocation. 
Since $i < 1$, the generalized mean inequality~(\ref{proof:NPhard_mean_ineq}) implies that $opt_{\cP_i}(t_M(x_{3DM}))$ is also upper bounded by $3$ and the equality holds when all individual data rates are equal to $3$, i.e.,
\begin{multline}\label{proof:NPhard_mean_eq}
opt_{\cP_i}(t_M(x_{3DM})) = opt_{\cP_1}(t_M(x_{3DM})) = 3 \\\iff \forall k\in\cK, \; R_k(\bp^*) = 3,
\end{multline}
where $\bp^*$ is an optimal power allocation of either $\cP_i$ or $\cP_1$ (this choice does not matter, as they are equal when~(\ref{proof:NPhard_mean_eq}) is satisfied). Finally, we derive equivalence~(\ref{proof:NPhard_final}) from~(\ref{proof:NPhard_2}) and~(\ref{proof:NPhard_mean_eq}), which proves that~${\cD_{i|M}}$ is strongly NP-hard.
\begin{equation}\label{proof:NPhard_final}
x_{3DM} \text{ has a matching} \iff opt_{\cP_i}(t_M(x_{3DM})) \geq 3.
\end{equation}
\end{IEEEproof}

As shown in Theorem~\ref{th:NPhard}, computing the optimal solution of~\ref{P} in the general case is intractable, unless P = NP. Nevertheless, we present in the next subsection some special cases in which~\ref{P} is solvable in polynomial time.

\subsection{Special Cases}\label{subsec:special_cases}
We highlight here four tractable special cases and discuss about possible polynomial time algorithms to solve them.

1) For a given subcarrier allocation $\cU_n$, $n\in\cN$, problem~\ref{P} reduces to a power control problem. In downlink, the sum-rate objective function with equal weights ${\bw =\{1/K,\ldots,1/K\}}$ can be rewritten as:
\begin{align}
& \cM_{1,\bw}\left(\bR\left(\bp\right)\right) = \sum_{n\in\cN}W_n\sum_{i=1}^{|\cU_n|}{R_{\pi_n(i)}^n} \nonumber\\
& = \sum_{n\in\cN}W_n\sum_{i=1}^{|\cU_n|}\log_2 \left(1+\frac{g_{\pi_n(i)}^n p_{\pi_n(i)}^n}{\sum_{j=i+1}^{|\cU_n|}{g_{\pi_n(i)}^n p_{\pi_n(j)}^n}+\eta^n_{\pi_n(i)}}\right) \nonumber\\
& = \sum_{n\in\cN}W_n\sum_{i=1}^{|\cU_n|}\log_2 \left(\frac{\sum_{j=i}^{|\cU_n|}{p_{\pi_n(j)}^n}+\eta^n_{\pi_n(i)}/g_{\pi_n(i)}^n}{\sum_{j=i+1}^{|\cU_n|}{p_{\pi_n(j)}^n}+\eta^n_{\pi_n(i)}/g_{\pi_n(i)}^n}\right) \nonumber\\
& = \sum_{n=1}^{N}W_n \left (\sum_{i=1}^{|\cU_n|-1} \log_2 (\alpha_i^n(\bp) ) + \log_2(\beta^n(\bp) ) \right). \label{concave_DL}
\end{align}
For $n\in\cN$ and $i<|\cU_n|$, $\alpha_i^n$ is obtained by combining the numerator of $R_{\pi_n(i+1)}^n$ and the denominator of $R_{\pi_n(i)}^n$, i.e.,
\begin{equation*}
\alpha_i^n(\bp) \triangleq \frac{\sum_{j=i+1}^{|\cU_n|}{p_{\pi_n(j)}^n}+\eta^n_{\pi_n(i+1)}/g_{\pi_n(i+1)}^n}{\sum_{j=i+1}^{|\cU_n|}{p_{\pi_n(j)}^n}+\eta^n_{\pi_n(i)}/g_{\pi_n(i)}^n},
\end{equation*}
and $\beta^n$ contains the numerator of $R_{1}^n$ and the denominator of $R_{|\cU_n|}^n$, i.e.,
\begin{equation*}
\beta^n(\bp) \triangleq \frac{\sum_{j=1}^{|\cU_n|}{p_{\pi_n(j)}^n}+\eta^n_{\pi_n(1)}/g_{\pi_n(1)}^n}{\eta^n_{\pi_n(|\cU_n|)}/g_{\pi_n(|\cU_n|)}^n}.
\end{equation*}

Assuming optimal decoding order~(\ref{SIC_DL}) is applied in downlink, we have $\eta_{\pi_n(i)}^n / g_{\pi_n(i)}^n \geq \eta_{\pi_n(i+1)}^n / g_{\pi_n(i+1)}^n$, for all $i<|\cU_n|$. It can be verified that $\alpha_i^n$ is a concave homographic function and $\beta^n$ is linear, therefore also concave. Thus, by composition with logarithms and summation, we derive that the objective function~(\ref{concave_DL}) is concave. In addition, the feasible set defined by $C1$ to $C4$ is a convex set. Therefore, given a fixed and arbitrarily chosen subcarrier allocation, the sum-rate maximization problem can be optimally solved using classical convex programming methods~\cite{boyd2004convex}. The same result applies to uplink transmissions with sum-rate rewritten as:
\begin{equation*}\label{concave_UL}
\cM_{1,\bw}\left(\bR\left(\bp\right)\right) = \sum_{n\in\cN}W_n \log_2 \left(\frac{\sum_{j=1}^{|\cU_n|}{g_{\pi_n(j)}^n p_{\pi_n(j)}^n}+\eta^n}{\eta^n}\right).  \\
\end{equation*}
In particular, if $M=K$, then all users can be multiplexed on all subcarriers. It directly follows that~\ref{P} is solvable by convex programming such as the projected gradient descent.

2) Without the above assumption, and if $K=1$, then the optimal data rate is given by the waterfilling power allocation~\cite{cover2012elements}.

3) In case there is only one subcarrier, i.e., $N=1$, a simple algorithm consists in sorting all users $k\in\cK$ by their SNR values without interference $g_k^1 p_k^1/\eta_k^1$, where $p_k^1 = \min{\{\bar{P}_k,\bar{p}_k^1\}}$ denotes the maximum power budget of user $k$. Then only the top-$M$ users are active with power $p_k^1$.

4) Finally, if the individual power constraint $C1$ is relaxed to a cellular power constraint, i.e., ${\sum_{k\in\cK}\sum_{n\in\cN}p_k^n \leq \bar{P}}$, then the sum-rate maximization problem becomes solvable in polynomial time according to~\cite{liu2014complexity1}. Furthermore, optimal strategies for all generalized mean utilities have been derived in~\cite{altman2008generalized} for OFDMA systems ($M=1$) subject to cellular power constraints.

\section{Conclusion}\label{sec:ccl}
In this paper, we develop a general framework to study the complexity of various resource allocation problems related to NOMA. The aim is to have a better understanding of these problems and to facilitate the design of resource allocation algorithms. In this framework, we prove that joint subcarrier and power allocation problems are strongly NP-hard by pseudo-polynomial reduction from the 3-dimensional matching (3DM). This result holds for any objective function which can be represented as a weighted generalized mean of order $i\leq 1$, e.g., weighted sum-rate, proportional fairness, harmonic mean and max-min fairness utilities. It is also valid for both downlink and uplink scenarios, as well as any number of multiplexed users per subcarrier. Furthermore, we present some tractable special cases which can be easily solved by polynomial time algorithms. 

\IEEEtriggeratref{3}
\bibliographystyle{IEEEtran}
\bibliography{IEEEabrv,reference}

\end{document}